\documentclass[a4paper,11pt]{article}
\usepackage{jheppub} 
\usepackage{lineno}
\pdfoutput=1

\usepackage{siunitx}
\usepackage{xcolor}
\usepackage{slashed}

\usepackage{amsmath,amssymb,mathtools}
\allowdisplaybreaks

\usepackage{float}
\usepackage{graphicx}
\usepackage{subcaption}
\usepackage[font=small,labelfont=bf]{caption}

\usepackage{feynmf}

\usepackage[normalem]{ulem}

\title{\boldmath Leptogenesis without on-shell right-handed neutrinos}

\author[a]{Simon Cléry}
\author[a]{Alejandro Ibarra}
\author[a]{Onur Yonar}
\affiliation[a]{Technical University of Munich, TUM School of Natural Sciences, Physics Department, 85748 Garching, Germany}

\emailAdd{simon.clery@tum.de, ibarra@tum.de, onur.yonar@tum.de}

\abstract{
We propose a novel mechanism for generating the baryon asymmetry of the Universe through leptogenesis in a scenario where the right-handed neutrinos are heavier than the maximal temperature of the Universe, and are never produced on-shell, neither by thermal nor by non-thermal mechanisms. We introduce a new scalar field, $\phi$, lighter than the right-handed neutrinos, that couples to the latter via a Yukawa coupling, so that it decays into two lepton doublets and two higgs doublets via off-shell right-handed neutrinos. Then, we derive the CP asymmetry arising from the interference between tree-level and loop diagrams in the four-body decay, and we show that the generated baryon asymmetry can reproduce the observed value both in a scenario where $\phi$ is responsible for the reheating of the Universe, and in a scenario where $\phi$ is a generic scalar that remains in thermal equilibrium with the plasma. 
}

\begin{document}
\unitlength=1mm
\maketitle
\flushbottom
\section{Introduction}

Leptogenesis is a remarkably simple and plausible mechanism to generate the cosmic quark-antiquark asymmetry~\cite{Fukugita:1986hr} (for a review, see \cite{Davidson:2008bu}).  In its simplest realization, a population of right-handed neutrinos is produced thermally or non-thermally in the very early Universe. Their subsequent decay out of equilibrium into Standard Model particles, with different rates between the CP transformed processes~\cite{Sakharov:1967dj}, generates a lepton asymmetry which is partially converted into a baryon asymmetry by sphaleron processes~\cite{Kuzmin:1985mm}.

The discovery of neutrino oscillations at the end of the 20th century \cite{Super-Kamiokande:1998kpq} provided an additional motivation for leptogenesis, as the necessary right-handed neutrino Majorana fields imply the existence of non-zero neutrino masses~\cite{Minkowski:1977sc,
Gell-Mann:1979vob,Yanagida:1979as}.
The measurement of small neutrino masses in turn implies that the lightest right-handed neutrino mass must be very large, in order to generate a sufficiently large CP asymmetry to ultimately generate the measured quark-antiquark asymmetry~\cite{Davidson:2002qv,
Hamaguchi:2001gw,Buchmuller:2002rq,Giudice:2003jh}. In turn, the maximal temperature of the Universe must be very large in order to populate the primeval plasma with right-handed neutrinos.

In this paper, we argue that a lepton asymmetry can be generated even if the right-handed neutrinos were never present in the primeval plasma, {\it i.e.} they were never produced neither thermally nor non-thermally.  Related leptogenesis scenarios where heavy neutral leptons are never produced on shell, but enter only virtually or through effective interactions, have been considered in the context of primordial lepton flavor oscillations during reheating \cite{Hamada:2016oft,Hamada:2018epb}, and recently for particle production from bubble collisions during first-order phase transitions \cite{Cataldi:2025nac}. Our mechanism instead relies on a direct CP asymmetry generation in four-body decay mediated by virtual right-handed neutrinos. We introduce a scalar field $\phi$ (which may play the role of the inflaton), lighter than the right-handed neutrino fields, and which decays  
into two lepton doublets and two Higgs doublets via its Yukawa interaction with (off-shell) right-handed neutrinos. We show that this decay can be C and CP violating and can generate an excess of antileptons over leptons in the early Universe, which eventually generates more quarks than antiquarks. 

The paper is organized as follows. In Section \ref{sec:CP_asymmetry} we calculate the CP asymmetry generated in the four-body decay of the scalar $\phi$ into two leptons and two Higgs doublets. In Section \ref{sec:inflaton} we identify the scalar $\phi$ with the inflaton, and we calculate the asymmetry generated during reheating. Instead, in Section \ref{sec:generic_scalar} we identify $\phi$ with a generic scalar in thermal equilibrium in the early Universe. Finally, in Section \ref{sec:conclusions} we present our conclusions. We also include Appendix \ref{app:CP} with details of the calculation of the CP asymmetry.

\section{CP asymmetry from the decay of a heavy scalar}
\label{sec:CP_asymmetry}
 We consider an extension of the SM field content by at least two Majorana right-handed neutrinos $N_i$, with masses $M_i$, ordered so that $N_1$ is the lightest eigenstate. In addition, we introduce  a complex scalar field $\phi$, with mass $M_\phi<M_i$. The relevant Lagrangian terms are,
\begin{equation}
\begin{aligned}
     - \mathcal{L}&\supset \dfrac{1}{2} M_i \overline{N_{i}^c}N_{i}  +\dfrac{1}{2}y_{ij}\overline{N_{i}^c}N_{j} \phi +(Y_\nu)_{\ell j}^\ast \overline L_\ell N_j  	\tilde H  +V(\phi) + {\rm h.c.} 
\end{aligned}
\end{equation}
with $L_\ell=(\nu_{\ell L},\ell_L)$ the lepton doublet ($\ell=e,\mu,\tau$), $\Tilde{H} = i \sigma_2 H^\ast$ with $H=(H^+ , H^0)^T$ is the conjugated Higgs doublet,  $y$ and $Y_\nu$ the Yukawa matrices, and $V(\phi)$ the scalar potential for $\phi$. 

If the maximum temperature of the Universe is larger than $M_1$, the Universe would be in general populated by right-handed neutrinos, produced either thermally via scatterings with particles from the Standard Model plasma or non-thermally. In this case, the out-of-equilibrium decay of the right-handed neutrino into $LH$ and $\overline L H^\dagger$, with slightly different rates due a one-loop generated CP asymmetry, generates a lepton-asymmetry which is ultimately transferred into a baryon asymmetry via sphaleron processes~\cite{Fukugita:1986hr}.

In this paper, we consider instead a scenario where the right-handed neutrino masses are larger than the maximum temperature of the Universe, and are never produced on-shell via thermal or non-thermal processes. 
If this is the case,  and since the right-handed neutrinos are never present in the Universe, there is no lepton asymmetry generated in their decays. On the other hand, a lepton asymmetry could be generated through the four-body decays $\phi\rightarrow L_\ell H L_{\ell'} H$ and its CP conjugated process, with the right-handed neutrinos off-shell. 

\begin{figure}[t!]
\centering
\fmfset{arrow_len}{3mm}
\fmfset{arrow_ang}{12}
\newcommand{\fcomma}{{,}\,}
%========================================================
% Row 1
%========================================================
\begin{minipage}[t]{0.78\textwidth}
\centering
\begin{fmfgraph*}(55,48)
    \fmfstraight
    \fmfleft{i}
    \fmfright{l1,h1,l2,h2}

    \fmf{scalar,label=$\phi$,label.side=left}{i,v0}
    \fmf{plain,label=$N_{i'}$,label.side=left}{v0,v1}
    \fmf{plain,label=$N_i$,label.side=right}{v0,v2}

    \fmf{scalar}{v1,l2}
    \fmf{fermion}{v1,h2}
    \fmf{fermion}{v2,l1}
    \fmf{scalar}{v2,h1}

    \fmfforce{(0.22w,0.50h)}{v0}
    \fmfforce{(0.53w,0.76h)}{v1}
    \fmfforce{(0.53w,0.24h)}{v2}

    \fmflabel{$L_{\ell}$}{l1}
    \fmflabel{$H$}{h1}
    \fmflabel{$L_{\ell'}$}{h2}
    \fmflabel{$H$}{l2}
\end{fmfgraph*}

{\small (a) Tree}
\end{minipage}

\vspace{1cm}

\begin{minipage}[t]{0.48\textwidth}
\centering
\begin{fmfgraph*}(55,48)
    \fmfstraight
    \fmfleft{i}
    \fmfright{h2,l2,h1,l1}

    \fmf{scalar,label=$\phi$,label.side=left}{i,v0}
    \fmf{plain,label=$N_{i'}$,label.side=left}{v0,v1}
    \fmf{plain,label=$N_{i}$,label.side=right}{v0,v2}

    % upper triangular vertex correction
    \fmf{fermion,label=$L_{\ell''}$,label.side=left}{d,v1}
    \fmf{scalar,label=$H$,label.side=right}{u,v1}
    \fmf{plain,label=$N_k$,label.side=left}{u,d}

    % final legs: now symmetric and almost straight from u,d
    \fmf{fermion}{u,l1}
    \fmf{scalar}{d,h1}

    % lower branch: symmetric pair from v2
    \fmf{fermion}{v2,h2}
    \fmf{scalar}{v2,l2}

    % geometry
    \fmfforce{(0.16w,0.50h)}{v0}
    \fmfforce{(0.42w,0.80h)}{v1}
    \fmfforce{(0.42w,0.30h)}{v2}
    \fmfforce{(0.68w,0.95h)}{u}
    \fmfforce{(0.68w,0.70h)}{d}

    % labels
    \fmflabel{$L_{\ell}$}{h2}
    \fmflabel{$H$}{l2}
    \fmflabel{$L_{\ell'}$}{h1}
    \fmflabel{$H$}{l1}
\end{fmfgraph*}

{\small (b) Vertex}
\end{minipage}
\hfill
\begin{minipage}[t]{0.48\textwidth}
\centering
\begin{fmfgraph*}(55,48)
    \fmfstraight
    \fmfleft{i}
    \fmfright{l1,h1,l2,h2}

    \fmf{scalar,label=$\phi$,label.side=left}{i,v0}
    \fmf{plain,label=$N_{i'}$,label.side=left}{v0,a}
    \fmf{plain,label=$N_i$,label.side=right}{v0,d}

    % self-energy loop on upper branch
    \fmf{plain,left,tension=0.35,
      label={$L_{\ell''}\fcomma \overline{L_{\ell''}}$},
      label.side=left}{a,b}
    \fmf{dashes,left,tension=0.35,
      label={$H\fcomma H^{\dag}$},
      label.side=left}{b,a}

    % corrected propagator goes to a dedicated decay vertex e
    \fmf{plain,label=$N_k$,label.side=left}{b,e}

    % THIS is the symmetric upper decay
    \fmf{fermion}{e,l2}
    \fmf{scalar}{e,h2}

    % lower branch: leave it however you like
    \fmf{fermion}{d,l1}
    \fmf{scalar}{d,h1}

    % geometry
    \fmfforce{(0.15w,0.50h)}{v0}
    \fmfforce{(0.36w,0.72h)}{a}
    \fmfforce{(0.58w,0.8h)}{b}
    \fmfforce{(0.8w,0.8h)}{e}   % center of the symmetric (alpha1,beta1) split
    \fmfforce{(0.42w,0.24h)}{d}

    % external labels
    \fmflabel{$L_{\ell}$}{l1}
    \fmflabel{$H$}{h1}
    \fmflabel{$L_{\ell'}$}{l2}
    \fmflabel{$H$}{h2}
\end{fmfgraph*}

{\small (c) Wave-function}
\end{minipage}

\vspace{1cm}

\begin{minipage}[t]{0.48\textwidth}
\centering
\begin{fmfgraph*}(55,48)
    \fmfstraight
    \fmfleft{i}
    \fmfright{l1,l2,h2,h1}

    \fmf{scalar,label=$\phi$,label.side=left}{i,v0}
    \fmf{plain,label=$N_{i'}$,label.side=left}{v0,a}
    \fmf{plain,label=$N_i$,label.side=right}{v0,b}

    \fmf{fermion}{a,h1}
    \fmf{scalar}{b,l1}

    \fmf{scalar,label=$H$,label.side=left}{a,c}
    \fmf{fermion,label=$L_{\ell''}$,label.side=right}{b,c}

    \fmf{plain,label=$N_k$,label.side=left}{c,d}
    \fmf{scalar}{d,h2}
    \fmf{fermion}{d,l2}

    \fmfforce{(0.15w,0.50h)}{v0}
    \fmfforce{(0.38w,0.73h)}{a}
    \fmfforce{(0.38w,0.27h)}{b}
    \fmfforce{(0.60w,0.50h)}{c}
    \fmfforce{(0.82w,0.50h)}{d}

    \fmflabel{$L_{\ell}$}{l2}
    \fmflabel{$L_{\ell'}$}{h1}
    \fmflabel{$H$}{l1}
    \fmflabel{$H$}{h2}
\end{fmfgraph*}

{\small (d) Scalar-contact topology}
\end{minipage}
\hfill
\begin{minipage}[t]{0.48\textwidth}
\centering
\begin{fmfgraph*}(55,48)
    \fmfstraight
    \fmfleft{i}
    \fmfright{l1,h1,h2,l2}

    \fmf{scalar,label=$\phi$,label.side=left}{i,v0}
    \fmf{plain,label=$N_{i'}$,label.side=left}{v0,a}
    \fmf{plain,label=$N_i$,label.side=right}{v0,b}

    \fmf{fermion}{a,l2}
    \fmf{fermion}{b,l1}

    \fmf{scalar,label=$H$,label.side=left}{a,c}
    \fmf{scalar,label=$H$,label.side=right}{b,c}

    \fmf{scalar}{c,h1}
    \fmf{scalar}{c,h2}

    \fmfforce{(0.17w,0.50h)}{v0}
    \fmfforce{(0.41w,0.76h)}{a}
    \fmfforce{(0.41w,0.24h)}{b}
    \fmfforce{(0.68w,0.50h)}{c}

    \fmfdot{c}

    \fmflabel{$L_{\ell}$}{l1}
    \fmflabel{$H$}{h1}
    \fmflabel{$H$}{h2}
    \fmflabel{$L_{\ell'}$}{l2}
\end{fmfgraph*}

{\small (e) Box}
\end{minipage}

\caption{Diagrams inducing the decay $\phi\rightarrow L_\ell H L_{\ell'} H, \overline{L_{\ell}} H^\dagger  \overline{L_{\ell'}} H^\dagger$ at tree level (top panel) and at the one loop level (middle and bottom panels). Here, $\ell$ and $\ell'$ denote leptonic flavors. The CP conjugated diagrams, as well as diagrams crossing the external legs are not shown for simplicity, although they are included in the calculation.}
\label{fig:all_diagrams}
\end{figure}
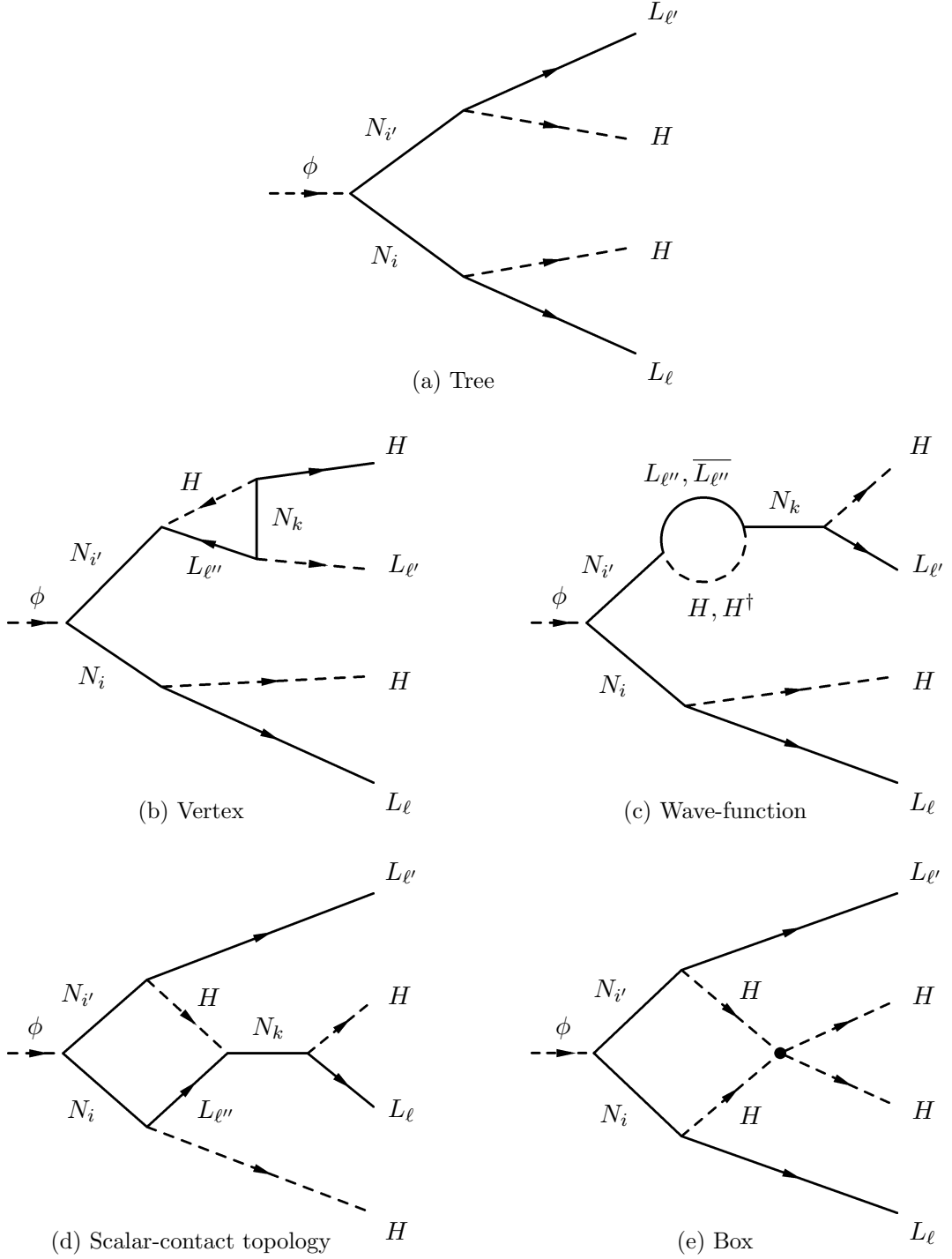

The CP asymmetry generated in the $1\rightarrow 4$ decay is defined as,
\begin{equation}
    \epsilon_{\text{CP}}=\dfrac{\Gamma(\phi \rightarrow L H L H)-\Gamma(\phi \rightarrow \overline{L} H^{\dag} \overline{L} H^{\dag})}{\Gamma(\phi \rightarrow L H L H)+\Gamma(\phi \rightarrow \overline{L} H^{\dag} \overline{L} H^{\dag})}\,
    ,
\end{equation}
where in the rates we have summed over all possible combinations of leptonic flavors. 
The interference between the tree-level and one-loop diagrams depicted in Fig.\ref{fig:all_diagrams} generates a CP asymmetry. Summing over all the lepton flavors we obtain  (for details, see Appendix \ref{app:CP}):\begin{align}
    \label{eq:eps-start-tot}
\epsilon_{\mathrm{tot}}
&\simeq - \frac{3 M_\phi^2}{320 \pi}\, \dfrac{\displaystyle{\sum_{i, i', q,q',k}} \dfrac{{\rm Im}\left[\mathcal{H}_{qi} y_{ii'}  \mathcal{H^*}_{i'k}  \mathcal{H}_{k q'} y_{q'q}^* \right]}{ M_k M_q^2 M_{q'} M_i M_{i'}}}{\displaystyle{\sum_{i,i', q,q'}\dfrac{\mathcal{H}_{qi}y_{ii'}\mathcal{H^*}_{i'q'}  y_{q'q}^*}{ M_q M_{q'} M_i M_{i'}}}}\,
,
\end{align}
where $\mathcal{H} \;\equiv\; Y_\nu^\dagger Y_\nu$. In the scenario in which the field $\phi$ couples just to the lightest right-handed neutrino, $N_{1}$, and assuming a strong hierarchy among the masses, we find 
\begin{equation}
\begin{aligned}
    \label{eq:eps-tot-leading}
\epsilon_{\mathrm{tot}}
&\simeq -\frac{3}{ 320\pi} \frac{M_\phi^2}{M_1^2} \sum_{k\neq1} \dfrac{M_1}{ M_k}\dfrac{{\rm Im}\left[\mathcal{H}^2_{k1} \right]}{ \mathcal{H}_{11}} \;.
\end{aligned}
\end{equation}

The CP asymmetry generated in the decays $\phi\rightarrow LH LH, \overline{ L} H^\dagger \overline{L} H^\dagger$ could be compared to the corresponding one generated in the decays of the lightest right-handed neutrino $N\rightarrow LH, \overline L H^\dagger$ \cite{Davidson:2002qv,Davidson:2008bu},
\begin{equation}
    \epsilon_\ell^{N_i}= \dfrac{1}{8 \pi}  \sum_{k\neq1} f\left(\dfrac{M_k^2}{M_1^2}\right) \dfrac{{\rm Im}\left[\mathcal{H}^2_{ki}\right]}{\mathcal{H}_{ii}}\simeq-\dfrac{3M_1}{16 \pi} \sum_{k\neq1}  \dfrac{1}{M_k}\dfrac{{\rm Im}\left[\mathcal{H}^2_{ki}\right]}{\mathcal{H}_{ii}}\, 
    ,
\end{equation}
where  $f(x)= \sqrt{x}(1-(1+x)\log((1+x)/x))$ and where in the last step we have taken the limit $M_1\ll M_k$. The CP asymmetry generated in the 4-body decay is a factor $M^2_\phi/(20 M^2_1)$ smaller than the one generated in the 2-body decay. Further, one can use the techniques of \cite{Davidson:2002qv} to obtain an upper limit on the CP asymmetry in the hierarchical right-handed neutrino mass scenario from requiring compatibility of the seesaw scenario with the measured neutrino masses. The result is,
\begin{equation}
    \begin{aligned}
        |\epsilon_{\mathrm{tot}}|\lesssim \frac{3}{320\pi}\dfrac{M_\phi^2}{M_1^2}\dfrac{M_1 (m_3-m_1)}{v^2}\, .
    \end{aligned}
\label{eq:epsilon}
\end{equation}

\begin{fmffile}{all_diagrams}

\end{fmffile}

\section{Baryon asymmetry from inflationary reheating}
\label{sec:inflaton}

In this section, we identify the field $\phi$ with the inflaton field. As a concrete realization of inflation, we consider the well-motivated Starobinsky model \cite{Starobinsky:1980te} for which the inflaton is a real scalar field with potential given by: 
\begin{equation}
V(\phi) = \frac{3}{4}M_\phi^2 M_P^2\left( 1- e^{-\sqrt{\frac{2}{3}}\frac{\phi}{M_P}}\right)^2\,
,
\label{eq:staro}
\end{equation}
where $M_\phi$ is the inflaton mass parameter and $M_P = 1/\sqrt{8 \pi G} \sim 2.4 \times 10^{18} ~\text{GeV}$ is the reduced Planck mass. The inflaton mass, $M_\phi$ is fixed by the amplitude of the scalar perturbations inferred from CMB measurements \cite{Planck:2018jri}, and for the Starobinsky potential, this gives $M_\phi \simeq 3 \times 10^{13}~\rm GeV$. 

The reheating occurs at the end of the slow-roll regime, as the inflaton scalar field oscillates about the minimum of its potential, located at $\langle \phi \rangle = 0$.  The end of inflation occurs when $\ddot a = 0$ which corresponds to an associated inflaton energy density $\rho_{\rm end} = \frac{3}{2}V(\phi_{\rm end})$. For the Starobinsky potential, $\rho_{\rm end} \simeq 0.175 M_\phi^2 M_P^2 \simeq \left(5.5\times10^{15}~\rm GeV\right)^4 $ \cite{Starobinsky:1980te}. In what follows, we work in the perturbative regime with $\phi\ll M_P$, hence the potential is approximated as
\begin{equation}
V(\phi\ll M_P)\simeq \frac{1}{2}M_\phi^2 \phi^2 \, .
\end{equation}
The inflaton then starts oscillating about a quadratic minimum, and the average equation of state of the Universe is thus given by $w_\phi = \frac{P_\phi}{\rho_\phi} \simeq 0$ during reheating.

In the previous section, we showed that the decay of a field $\phi$ into off-shell right-handed neutrinos can generate a lepton asymmetry. If $\phi$ is the inflaton field, we then expect a lepton asymmetry to be generated during the reheating era, which is transferred into a baryon asymmetry by the sphalerons.

To track the production and evolution of the lepton asymmetry, we write down the Boltzmann equations governing the evolution of the energy density of the inflation (which evolves as matter), and the total energy density of all Standard Model fields (which evolve as radiation),
\begin{equation}
\begin{aligned}
&\dot{\rho}_\phi + 3 H \rho_\phi = - \Gamma_\phi \rho_\phi\,, \\
&\dot{\rho}_R + 4 H \rho_R = \Gamma_\phi \rho_\phi\,.
 \end{aligned}
 \label{eq:BE}
\end{equation}
Here, $H\equiv \sqrt{\rho_\phi + \rho_R}/\sqrt{3}M_P$ is the Hubble rate and $\Gamma_\phi$ is the total inflaton decay rate, which we assume approximately constant~\footnote{One should note that for anharmonic inflaton potential near the minimum, the scalar field develops a time-dependent effective mass, rendering any interaction rate of the inflaton time-dependent.}. Then the first equation can be solved approximately giving,
\begin{equation}
    \rho_\phi(a) \simeq \rho_{\rm end}\left( \frac{a_{\rm end}}{a}\right)^3\;,
    \label{eq:rhophi_sol}
\end{equation}
with $a_{\rm end}<a<a_{\rm RH}$ the scale factor between the end of inflation and the end of the reheating. Substituting Eq.~(\ref{eq:rhophi_sol}) into the Boltzmann equation for $\rho_R(a)$, one can readily obtain the evolution of the radiation density during reheating that reads \cite{Garcia:2020wiy},
\begin{equation}
    \rho_R(a) \simeq \frac{2\sqrt{3}}{5} \Gamma_\phi \,\rho_{\rm end}^{1/2} \, M_P \left(\frac{a_{\rm end}}{a}\right)^4\left[ \left(\frac{a}{a_{\rm end}}\right)^{5/2}-1\right] \;.
\end{equation}

We assume an instantaneous thermalization of the radiation bath during the reheating era, and define a time-dependent temperature $T(a)\equiv \left(30\rho_R(a)/\pi^2g_{\star} \right)^{1/4}$. The reheating process ends when $H(a)\equiv H_{\rm rh}\simeq \Gamma_\phi$ or equivalently when the radiation energy density satisfies $\rho_R(a_{\rm RH}) = \rho_\phi(a_{\rm RH})$. At that time, we can define the reheating temperature in the SM plasma as
\begin{align}
T_{\rm RH} = \left(\frac{30\,\rho_R(a_{\rm RH})}{\pi^2 g_{\star}}\right)^{1/4}\,
,
\end{align}
where the scale factor at reheating time can be computed approximately as
\begin{equation}
        a_{\rm RH} \simeq a_{\rm end} \left( \dfrac{5 H_{\rm end}}{2\Gamma_\phi} \right)^{2/3}
        \label{eq:a_RH}\, .
\end{equation}

It follows from this equation that the SM radiation density rapidly reaches a maximum value at $a=a_{\rm max} \simeq 1.5\, a_{\rm end}$, associated to a maximum temperature $T_{\rm max}$ given by \cite{Garcia:2020wiy}
\begin{equation}
    \rho_{R}(a_{\rm max}) \simeq 0.24\times \Gamma_\phi\, \rho_{\rm end}^{1/2}\, M_P \implies T_{\rm max} \simeq \left(\frac{0.7\times \Gamma_\phi\, \rho_{\rm end}^{1/2}\, M_P }{ g_{\star}}\right)^{1/4}\;.
\end{equation}
After reaching this maximum, the energy density in the bath scales as $\rho_R(a\gg a_{\rm end})\propto a^{-3/2}$, or equivalently, as temperature $T\propto a^{-3/8}$, which is not isentropic, indicating the injection of entropy from inflaton decays. One can compute from such an evolution the reheating temperature as
\begin{equation}
    T_{\rm RH} = \left(\frac{30\,}{\pi^2 g_{\star}}\right)^{1/4}\left(\frac{12}{25}\right)^{1/4}\sqrt{\Gamma_\phi M_P}\;,
    \label{eq:T_rh}
\end{equation}
from where one can readily calculate the evolution of the entropy density using,
\begin{equation}
    s(a) \equiv \frac{2\pi^2 g_\ast}{45}T^3(a)\;.
\end{equation}
Unless explicitly specified, we take the number of relativistic degrees of freedom $g_\ast = 106.75$.

Let us now discuss the evolution of the lepton asymmetry generated in the inflaton decay. In the following equation, we keep the inverse-decay contribution formally in the washout term, but we will work in the regime where thermal inverse production of on-shell $\phi$ particles is Boltzmann suppressed. This is ensured when the temperature of the thermal bath relevant for
these processes is below $M_\phi$, as verified below in the
viable region of parameter space. The Boltzmann equation takes the form, 

\begin{equation}
        \dfrac{dY_\Delta}{da} = +\frac{1}{a H(a)}\epsilon_\Delta \Gamma_\phi Y_\phi(a) -W(a)Y_\Delta(a)\,
        ,
        \label{eq:BE_asym}
\end{equation}
where $Y_\phi(a)=\rho_\phi(a)/\left(M_\phi s(a)\right)$ and $W(a)$ parametrizes the washout from both inverse decays (ID) and  $\Delta L =2$ scatterings,
\begin{equation}
    W(a) \equiv W_{\Delta L=2}(a) + W_{\rm ID}(a)\, 
    ,
\end{equation}
with
\begin{eqnarray}
    W_{\Delta L =2}(a) &\equiv& \frac{\gamma_{\Delta L =2}(a)}{aH(a)s(a) Y_L^{\rm eq}(a)} = \frac{\Gamma_{\Delta L =2}}{aH(a)},\\
    W_{\rm ID}(a) &\equiv& \frac{\Gamma_{\rm ID}}{aH(a)}\, 
    .
\end{eqnarray}

In the scenario considered in this work, the right-handed neutrinos are heavier than the temperature of the thermal plasma and are never present in the Universe, therefore the $\Delta L=1$ inverse decay process $L H\rightarrow N$ that dominates the washout in the standard leptogenesis scenarios is not present here. Moreover, in the viable region of parameter space considered below, the inflaton mass $M_\Phi$ is larger than the temperature of the thermal plasma during reheating, so that \(\Gamma_{\rm ID}\simeq0\).

This scenario however, predicts potentially relevant  $\Delta L =2 $ wash-out scattering processes $L H \leftrightarrow \overline{L} H^\dagger$ and $L L\leftrightarrow H^\dagger H^\dagger$ mediated by off-shell heavy right-handed neutrinos. We estimate the rate for $\Delta L=2$ scatterings to be 
\begin{equation}
    \Gamma_{\Delta L=2}^{2 \leftrightarrow 2} \sim  O(10^{-4})|Y_\nu|^4 \frac{T^3}{M_1^2}\sim
    O(10^{-4})\frac{T^3 m_3^2}{v^4}\;,
\end{equation}
where the numerical prefactor is an order-of-magnitude estimate obtained from thermal averaging \cite{Buchmuller:2002rq,Buchmuller:2004nz}, and we have used in the last step $m_3 \sim v^2 |Y_\nu|^2/M_1 $. This relation should be understood as a benchmark motivated by seesaw
realizations in which the atmospheric neutrino mass scale is dominantly generated by the lightest right-handed neutrino. We do not assume a specific flavor model in what follows.

The washout rate from $\Delta L =2$ scatterings scales with $T^3$, and therefore is most efficient during reheating, when $T=T_{\rm max}$, 
while the Hubble rate in this epoch scales as $H\sim \rho_\phi^{1/2}/M_P$. Therefore, the $\Delta L=2$ washout processes can therefore be neglected when $\Gamma_{\Delta L=2}/H(T=T_{\rm max})\lesssim 1$, which gives
\begin{equation}
    T_{\rm max} \lesssim 4.7\times 10^{15}\, {\rm GeV} \left( \frac{0.05\, {\rm eV}}{m_3}\right)^{2/3}\left(\frac{\rho_{\rm end}^{1/4}}{5.5\times 10^{15}\, \rm GeV}\right)^{2/3}\;,
\end{equation}
which is fulfilled in a large class of inflationary scenarios. On the other hand, it should be noted that the Hubble rate redshifts as $H\propto a^{-3/2}$, while $\Gamma_{\Delta L =2}\propto a^{-9/8}$. Therefore, the washout might also become efficient at the end of reheating. Requiring instead that $\Gamma_{\Delta L=2}/H(T=T_{\rm RH})\lesssim 1$, and using $H(T=T_{\rm RH})\sim \sqrt{g_*}\, T_{\rm RH}^2/M_P$, one obtains the following upper bound on the reheating temperature
\begin{equation}
    T_{\rm RH} \lesssim 2.1\times 10^{16}\, {\rm GeV} \left(\frac{0.05\, {\rm eV}}{m_3}\right)^2 \,
    ,
\end{equation}
which is in fact less stringent than the condition on maximum temperature. 

Finally, the inflaton coupling to heavy right-handed neutrino might also induce $\Delta L= 2$ processes, such as $\phi\,LL\rightarrow H^\dagger H^\dagger$, from the effective operator,
 \begin{equation}
 -\mathcal{L} \supset \mathcal{C}_6\, \frac{\phi (LH)(LH)}{M_1^2}\,
 ,
 \end{equation}
integrating out the heavy right-handed neutrino. We estimate the rate for this $3\leftrightarrow2$ process to be 
\begin{equation}
    \Gamma_{\Delta L =2}^{3\rightarrow2} \sim \xi{\cal C}^2_6 \frac{T^6}{M_1^4 E}\,
    ,
\end{equation}
where $E$ is the typical energy transferred in the process and $\xi\ll 1$ is some numerical factor due to phase space integrals. Since $T\ll M_\phi$ during reheating, and $ T \lesssim  E \lesssim M_\phi $, one obtains
\begin{equation}
    \Gamma_{\Delta L =2}^{3\rightarrow2} \lesssim \xi\,{\cal C}^2_6 \frac{T^6}{M_1^4 M_\phi}\,
    .
\end{equation}
Therefore, comparing with the Hubble expansion rate during reheating, one obtains an upper limit on the maximal temperature of the Universe so that the wash-out of the lepton asymmetry induced by $\Delta L=2$ processes involving the inflation field are negligible. We obtain
\begin{eqnarray}
    T_{\rm max}
    &\lesssim& 10^{13}\, {\rm GeV}\, \xi^{-1/6}\mathcal{C}_6^{-1/3} \left(\frac{M_1}{M_\phi}\right)^{2/3}\left(\frac{M_\phi}{10^{13}\, \rm GeV}\right)^{5/6}\left(\frac{\rho_{\rm end}^{1/4}}{5.5\times 10^{15}\, \rm GeV}\right)^{1/3}.
\end{eqnarray}
Since  $\xi \ll 1$, $M_1>M_\phi$, this upper bound on the maximum temperature is typically larger than $M_\phi$. Thus, in our model of reheating through perturbative decays of the inflaton, we can safely neglect the $3\rightarrow2$ via $\phi\, LL \rightarrow H^\dagger H^\dagger$, in the evolution and washout of the lepton asymmetry sourced by the inflaton decays. 

Since the washout is in general negligible, the Boltzmann equation for the asymmetry, Eq.~(\ref{eq:BE_asym}) can be analytically solved, the result being
\begin{equation}
    Y_\Delta(a) \simeq 0.5\times  \epsilon_\Delta \frac{\Gamma_\phi^{1/4}M_P^{1/4}\rho_{\rm end}^{1/8}}{M_\phi}\frac{\left[\left(\frac{a}{a_{\rm end}}\right)^{3/2}-1\right]}{\left[\left(\frac{a}{a_{\rm end}}\right)^{5/2}-1\right]^{3/4}}\,
    ,
    \label{eq:asym_yield}
\end{equation} 
which behaves as $Y_\Delta(a\gg a_{\rm end})\propto a^{-3/8}$ until it is frozen at $a=a_{\rm RH}$ when inflaton energy density becomes exponentially suppressed. From the asymmetry in the number density of leptons and antileptons, it is straightforward to calculate the baryon asymmetry at the time of freeze out,
\begin{equation}
    Y_B\big|_{a=a_{\rm RH}} = c_s Y_\Delta \big|_{a=a_{\rm RH}}= \frac{28}{79}Y_\Delta \big|_{a=a_{\rm RH}}\,
    ,
\end{equation}
where the prefactor 28/79 accounts for the conversion of the lepton asymmetry into a baryon asymmetry via the electroweak sphalerons \cite{Khlebnikov:1988sr,Buchmuller:2004nz}.  
Replacing $a_{\rm RH}/a_{\rm end}\rightarrow \left(\rho_{\rm end}/\rho_{\rm RH}\right)^{1/3}$ in Eq.(\ref{eq:asym_yield}), and using Eq.(\ref{eq:T_rh}), we obtain,
\begin{equation}
    Y_B \simeq 0.2\times g_\ast^{-1/4}\, \epsilon_\Delta \frac{\Gamma_\phi^{1/2}M_P^{1/2}}{M_\phi}\,
    ,
    \label{eq:Y_B-general}
\end{equation}
which depends on the phenomenological parameters $(\Gamma_\phi, \,\epsilon_\Delta)$, for the scenario where the inflaton $\phi$ is coupled exclusively to  RH neutrinos. In such a scenario we have,
\begin{eqnarray}
    \epsilon_\Delta &\simeq&  2.5 \times 10^{-5}  \left(\dfrac{M_\phi}{M_1} \right) \left(\dfrac{M_\phi}{\SI{e13}{\giga\electronvolt}}\right)\,
    ,
    \\
    \Gamma_\phi &\simeq& \SI{11.4}{\giga\electronvolt}\,  y^2 \left(\dfrac{M_\phi}{M_1}\right)^2\left(\dfrac{M_\phi}{\SI{e13}{\giga\electronvolt}}\right)^3 \,
    , \\
    T_{\rm RH} &\simeq& 1.8\times 10^{9}\, {\rm GeV} \,  y\left(\frac{M_\phi}{M_1}\right) \left(\dfrac{M_\phi}{\SI{e13}{\giga\electronvolt}}\right)^{3/2} \, ,
    \label{Eq:decay_width}
\end{eqnarray}
where we have assumed that the CP asymmetry generated in the decay is close to its theoretical upper limit Eq.~(\ref{eq:epsilon}), and we have adopted $m_3=0.05$ eV for the largest active neutrino mass.

Replacing these values in Eq.~(\ref{eq:Y_B-general}) we obtain
\begin{equation}
    Y_B \simeq 10^{-10}   \left(\frac{y}{5\times 10^{-2}}\right) \left(\frac{M_\phi}{M_1}\right)^2 \left(\dfrac{M_\phi}{\SI{e13}{\giga\electronvolt}}\right)^{3/2}    \, ,
\end{equation}
which is in the ballpark of the measured value for a moderate hierarchy of masses $M_1/M_\phi\lesssim O(10)$ (when the Yukawa couplings remain perturbative $y\lesssim \sqrt{4\pi}$). 
Further, the model predicts a relation between the reheating temperature and the baryon yield given by,
\begin{equation}
T_{\rm RH}\simeq 8.8\times 10^7\,{\rm GeV}\left(\frac{M_1}{M_\phi}\right) \left(\frac{Y_B}{10^{-10}}\right)\;.
\end{equation}

\begin{figure}
    \centering
\includegraphics[width=0.49\textwidth]{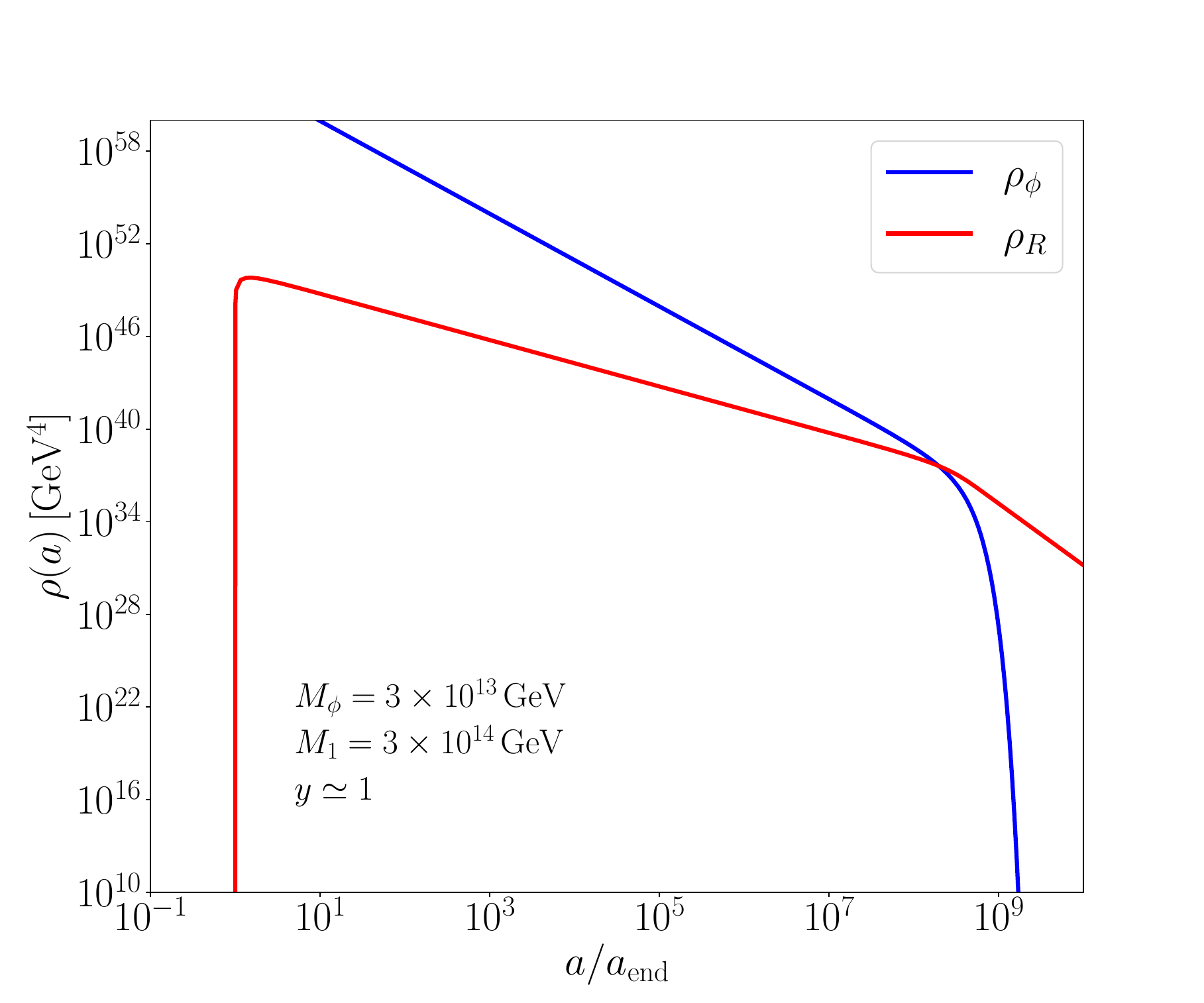}
\includegraphics[width=0.49\textwidth]{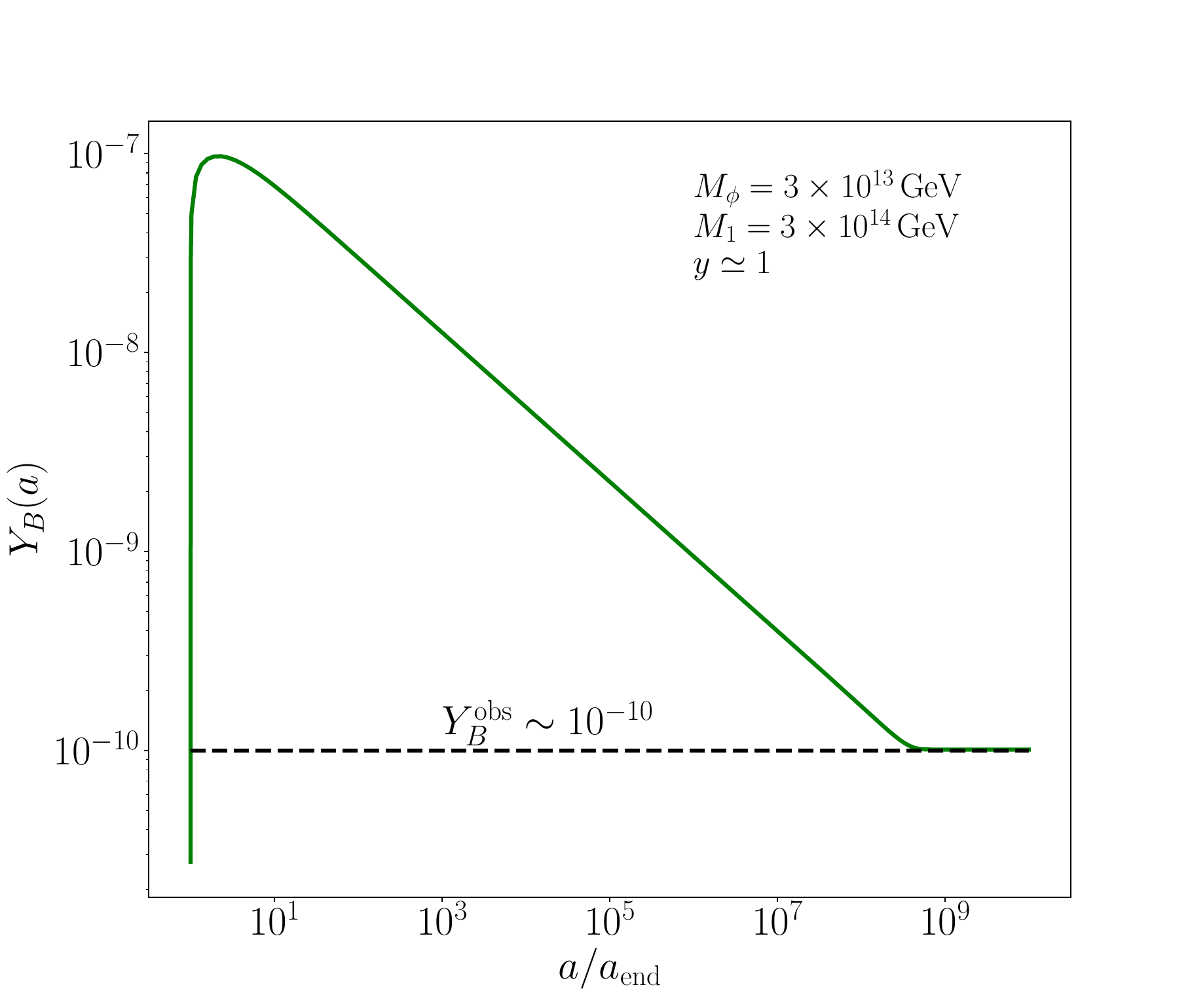}
    \caption{Evolution during reheating of the inflaton and radiation densities (left panel), as well as of the baryon asymmetry (right panel), for $M_\phi=3\times 10^{13}$ GeV, $M_1=3\times 10^{14}$ GeV and $y\simeq1$, assuming that the upper bound on the CP asymmetry Eq.~(\ref{eq:epsilon}) is saturated.}
    \label{fig:leptogenesis_reheating}
\end{figure}

We show in the left panel of Fig.~\ref{fig:leptogenesis_reheating} the evolution of the inflaton and radiation densities during reheating for the specific choice of parameters $M_\phi=3\times 10^{13}$ GeV, $M_1=3\times 10^{14}$ GeV and $y=1$. At very early times, the energy content of the Universe is dominated by the inflaton. However, the  inflaton decays  $\phi\rightarrow LH LH$ fill the Universe with radiation, which dominates the total energy content at $a/a_{\rm end}\sim 2\times 10^8$, which marks the end of reheating, $a_{\rm RH}$. The right panel shows the evolution of the baryon asymmetry, where we have assumed that the upper bound on the CP asymmetry Eq.~(\ref{eq:epsilon}) is saturated. As apparent from the plot, the same decay $\phi\rightarrow LH LH$ that reheats the Universe generates also a baryon asymmetry, which is diluted by the efficient entropy production during reheating,  until the end of reheating, when the Universe reaches thermal equilibrium and the total entropy remains constant. 

\section{Generic massive scalar in thermal equilibrium}
\label{sec:generic_scalar}

Let us now consider the scenario where the real scalar $\phi$ is not responsible for cosmic inflation and is one of the constituents of the thermal bath, in equilibrium with the SM particles.  The relevant $\phi$-dependent terms in the Lagrangian are
\begin{equation}
-\mathcal L \supset
\frac12 M_\phi^2\phi^2
+
\frac12\lambda_{\phi H}\phi^2 |H|^2
+\frac12 y_{ij} \overline{N_i^c} N_j \phi\; ,
\end{equation} 
and we neglect other possible interactions of $\phi$.
The renormalizable coupling to the Higgs induces $2\leftrightarrow 2$ scattering processes in the bath, which are expected to maintain the thermal equilibrium of the scalar field $\phi$ efficiently, until it becomes non-relativistic around $T\lesssim M_\phi$. Above electroweak symmetry breaking, the total thermally averaged cross section for $\phi\phi \leftrightarrow HH^\dagger$ is estimated as \cite{Gondolo:1990dk},
\begin{equation}
\langle \sigma_{2\leftrightarrow 2} v\rangle_T
\simeq
\frac{\lambda_{\phi H}^2}{16\pi \left(M_\phi^2 + cT^2\right)}\, ,
\end{equation}
where the coefficient $c\simeq4$ obtained from the thermal average allows to interpolate between relativistic to non-relativistic regime.

The temperature below which the scalar field is maintained in thermal equilibrium is obtained from requiring that the rate of $2\leftrightarrow 2$ scatterings is larger than the expansion rate $ n_{\rm eq}^\phi(T)\langle \sigma_{2\leftrightarrow2} v\rangle_T \gtrsim H(T)$, where $n_{\rm eq}^{\phi}(T)$ is the equilibrium number density for the scalar field, and is given by
\begin{equation}
    T_{2\leftrightarrow 2} \simeq 4.3 \times 10^{14}\, \rm GeV \times\lambda_{\phi H}^2\, ,
\end{equation}
where we have used the relativistic limit, $T\gg M_\phi$, in the thermal cross-section. 

For $T\lesssim T_{2 \leftrightarrow 2}$, the scalar $\phi$ is therefore in thermal equilibrium and its yield is  described by the following Boltzmann equation:
\begin{eqnarray}
     \dfrac{dY_\phi}{dz} &=& -\frac{1}{z H(z)}\frac{K_1(z)}{K_2(z)}\Gamma_\phi\left( Y_\phi(z) - Y_\phi^{\rm eq}(z)\right) -\frac{s\langle \sigma_{2\leftrightarrow2} v\rangle_T}{zH(z)}\left( Y_\phi^2(z) - \left(Y_\phi^{\rm eq}(z)\right)^2\right)
\end{eqnarray} 
which is now expressed in terms of $z\equiv M_\phi/T$, and where we have included the effect of the inverse reactions in the terms proportional to $Y_\phi^{\rm eq} = \frac{45}{4\pi^4g_\ast} z^2 K_2(z)$. 

\begin{figure}
    \centering
    \includegraphics[width=0.7\linewidth]{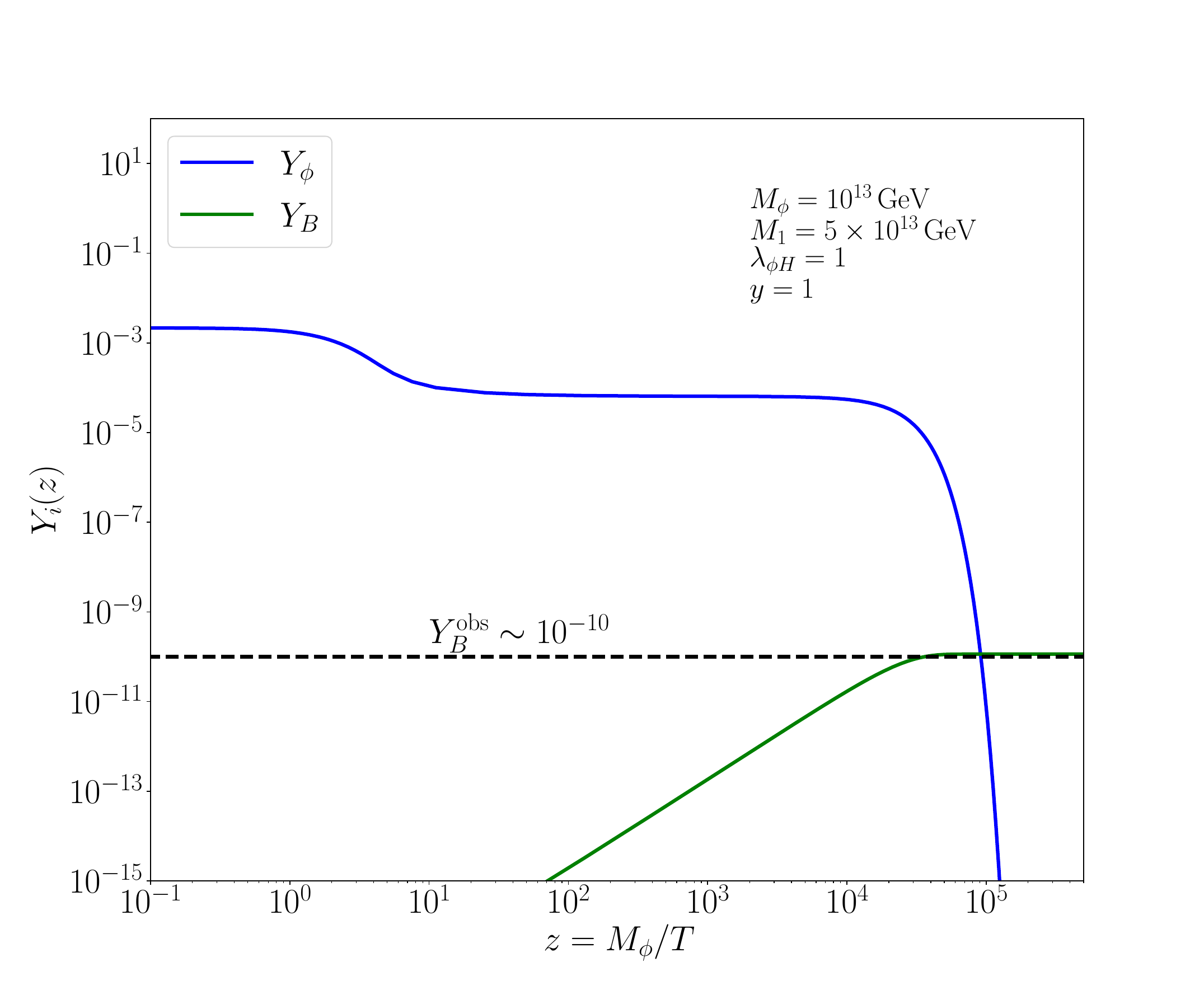}
    \caption{Numerical solutions of the thermal Boltzmann equations for a decaying massive scalar field. The baryon yield is frozen deeply in the non-relativistic regime as $\Gamma_\phi \ll M_\phi$, and matches the observed one for $M_\phi =  10^{13}\, \rm GeV$, $M_1 =   5 \times 10^{13}\, \rm GeV$, $y=1$, $\lambda_{\phi H}=1$.}
    \label{fig:leptogenesis_thermal}
\end{figure}

Around the time at which the scalar field becomes non-relativistic, it freezes-out from the bath with a relic yield~\cite{Kolb:1990vq},
\begin{equation}
    Y_{\rm FO} \simeq \frac{x_{\rm FO}}{\sqrt{g_*}\,M_PM_\phi\langle \sigma_{2\leftrightarrow2}  v\rangle_T} \simeq \frac{16\pi\, x_{\rm FO}  M_\phi}{\sqrt{g_*}\,M_P \lambda_{\phi H}^2} \,
    ,
\end{equation}
where we considered the non-relativistic limit, $T\ll M_\phi$, for the thermal cross section and introduce $x_{\rm FO}\equiv M_\phi/T_{\rm FO}$ as the redshift at freeze-out.

The Boltzmann equation for the lepton asymmetry generated in the $\phi\rightarrow LH LH$ decay, on the other hand, reads,
\begin{eqnarray} \label{eq:BE-asymmetry}
\dfrac{dY_\Delta}{dz} &=& \frac{1}{z H(z)}\frac{K_1(z)}{K_2(z)}\epsilon_\Delta \Gamma_\phi\left( Y_\phi(z) - Y_\phi^{\rm eq}(z)\right) -W(z)Y_\Delta(z)\,
,
\end{eqnarray} 
which includes a washout term. However,   similarly to the scenario described in Section \ref{sec:inflaton}, the washout in our scenario is in general negligible. First, we note that the decay of $\phi$ is rather slow, due to the large multiplicity of the final state. We estimate the decay temperature to be,
\begin{equation}
    T_D \simeq 9\times 10^{7}\, {\rm GeV} \times y\,\left(\frac{M_\phi}{M_1}\right)\left(\frac{M_\phi}{10^{12}\, \rm GeV}\right)^{3/2}\,
    .
\end{equation}
At this temperature, the Hubble expansion rate is much larger than the rate of inverse decays, thus making the washout due to the latter completely negligible.

Second, and using the same argument as in Section \ref{sec:inflaton}, the  $\Delta L =2$ washout processes induced by the Weinberg operator via $2\leftrightarrow2$ scatterings can be safely neglected around $T\sim T_D$ as long as $T_D \ll 10^{15}\,\rm GeV$. Lastly, the washout induced by $2\leftrightarrow 3$ processes induced by the  $\Delta L=2$ effective operator $\mathcal{O}_6\sim \mathcal{C}_6\, \phi (LH)(LH)/M_1^2$ can also be neglected at the temperature $T_D$, when most of the asymmetry is generated, and below. The interaction rate is given by
\begin{equation}
\Gamma_{\Delta L=2}^{3\leftrightarrow2} \equiv \frac{\gamma_{\Delta L=2}^{3\leftrightarrow2}}{n_\phi}= \frac{\gamma_{\Delta L =2}^{3\leftrightarrow2}(z)}{s(z) Y_\phi(z)}\,
,
\end{equation}
where 
\begin{equation}
\gamma_{\Delta L=2}^{3\leftrightarrow2}
\sim \kappa y^2 \left(\frac{M_1 m_3}{v^2}\right)^2 \frac{T^8}{M_1^4}\, .
\end{equation}
Therefore, the temperature at which the $2\leftrightarrow 3$ rate becomes smaller than the Hubble rate is,
\begin{equation}
T_{2\leftrightarrow3} \simeq 2.6 \times 10^{12}\,\mathrm{GeV}\,
\kappa^{-1/3} y^{-2/3}
\left(\frac{M_1}{M_\phi}\right)^{2/3}
\left(\frac{M_\phi}{10^{12}\,\mathrm{GeV}}\right)^{2/3}.
\end{equation}
Since $\kappa \ll 1$ and $M_1 > M_\phi$, one finds $T_{2\leftrightarrow3} \gg M_\phi$. Therefore, these processes are suppressed at $T \sim M_\phi$ and become negligible at $T \sim T_D \ll M_\phi$.

Neglecting the washout in Eq.~\eqref{eq:BE-asymmetry} one can estimate the produced lepton and baryon asymmetries from the yield of $\phi$ when their decays are completed, which occurs at $T\lesssim T_D$,
\begin{equation}
    Y_\Delta \sim \epsilon_\Delta Y_{\rm FO} \, \implies Y_B = c_s \epsilon_\Delta Y_{\rm FO} = \frac{28}{79}\epsilon_\Delta Y_{\rm FO}\,
    ,
\end{equation}
as illustrated in Figure \ref{fig:leptogenesis_thermal}.
This shows that the observed Baryon asymmetry is generated as long as $T_D \gtrsim T_{\rm EW}$ (in order for electroweak sphalerons to be in equilibrium before the last decays of the scalar field), and if the asymmetry parameter satisfies,\footnote{More generally, the above relation should be interpreted as a requirement on the relic abundance of the scalar field prior to its decay. While in the present realization this abundance is generated through thermal freeze-out via the Higgs portal interaction, any alternative mechanism producing a comparable non-relativistic abundance of $\phi$ before decay would lead to a similar baryon asymmetry. We leave the exploration of such alternative cosmological histories to future work.}
\begin{equation}
    \epsilon_\Delta \simeq \frac{79}{28}\frac{Y_B^{\rm obs}}{Y_{\rm FO}}\simeq \frac{3\times 10^{-10}} {Y_{\rm FO}}\,
    .
\end{equation}
Replacing the freeze-out yield into this constraint on the asymmetry parameter for obtaining the observed baryon asymmetry, one gets,
\begin{equation}
    \epsilon_\Delta \simeq 1.4\times 10^{8}\times \frac{\, \lambda_{\phi H}^2}{x_{\rm FO}M_\phi} \, .
\end{equation}

Assuming that the CP asymmetry in the decay is close to its maximal value Eq.(\ref{eq:epsilon}), one obtains that successful leptogenesis requires,
\begin{equation}\label{eq:lambda_exp}
    \lambda_{\phi H} \simeq 0.13\times\,x_\text{FO}^{1/2}\ \, \left(\frac{M_\phi}{M_1}\right)^{1/2}\left(\frac{M_\phi}{10^{12}\, \rm GeV}\right)\,
    .
\end{equation}

It should be noted that in our minimal set-up the heavy scalar decouples very early, while being non-relativistic. Therefore, the Universe could become matter dominated and the late decays of $\phi$ could dilute the asymmetry generated.  Such an early matter-dominated phase could also leave characteristic imprints on primordial stochastic gravitational wave backgrounds~\cite{Seto:2003kc, DEramo:2019tit, Pearce:2023kxp}. In the present work, however, we restrict ourselves to the radiation-dominated regime and leave the early-matter-dominated case and its gravitational-wave signatures to future work. To maintain the Universe dominated by radiation at early times, as assumed in our analysis, we require the decays to occur before the equality $\rho_\phi = \rho_R$ , which occurs at the temperature
\begin{equation}
    T_{\rm eq} \simeq 2.7\times10^{6}\, {\rm GeV}\, x_{\rm FO}\, \lambda_{\phi H}^{-2} \left(\frac{M_\phi}{10^{12}\,\rm GeV}\right)^2\,
    ,
\end{equation}
where we have used that $\rho_\phi = M_\phi Y_{\rm FO}s(T)$. Requiring now that $T_{\rm eq}<T_D$, we obtain the lower limit on the Yukawa coupling 
\begin{equation}\label{eq:y_limit_edm}
    y \gtrsim 3\times 10^{-2} \, x_{\rm FO}\, \lambda_{\phi H}^{-2} \left(\frac{M_\phi}{10^{12}\, \rm GeV}\right)^{1/2} \left(\frac{M_1}{M_\phi}\right) \,
    .
\end{equation}
Finally, replacing  Eq.~\eqref{eq:lambda_exp} into Eq.~\eqref{eq:y_limit_edm} we obtain
\begin{equation}
     y \gtrsim 0.16
     \left(\frac{M_1}{M_\phi}\right)^2
     \left(\frac{5\times10^{12}\, \rm GeV}{M_\phi}\right)^{3/2}\,
     ,
     \label{eq:EMD_constraint}
\end{equation}
which is required to ensure that the early Universe is always radiation dominated, and that the baryon asymmetry generated by the decay $\phi\rightarrow LH LH$ is in agreement with observations. In turn, this limit translates into a lower bound on the temperature at which the asymmetry is produced given by,
\begin{equation}
    T_D \gtrsim 1.6\times 10^{8} \,{\rm GeV} \left(\frac{M_1}{M_\phi}\right)\, .
\end{equation}
Interestingly, this lower limit does not depend on the Higgs portal coupling, and reflects the compensation between the abundance of the scalar field and the CP asymmetry required per decay: smaller values of $\lambda_{\phi H}$ lead to a larger scalar abundance, thereby reducing the asymmetry needed from each individual decay. 

\begin{figure}
    \centering
    \includegraphics[width=0.7\linewidth]{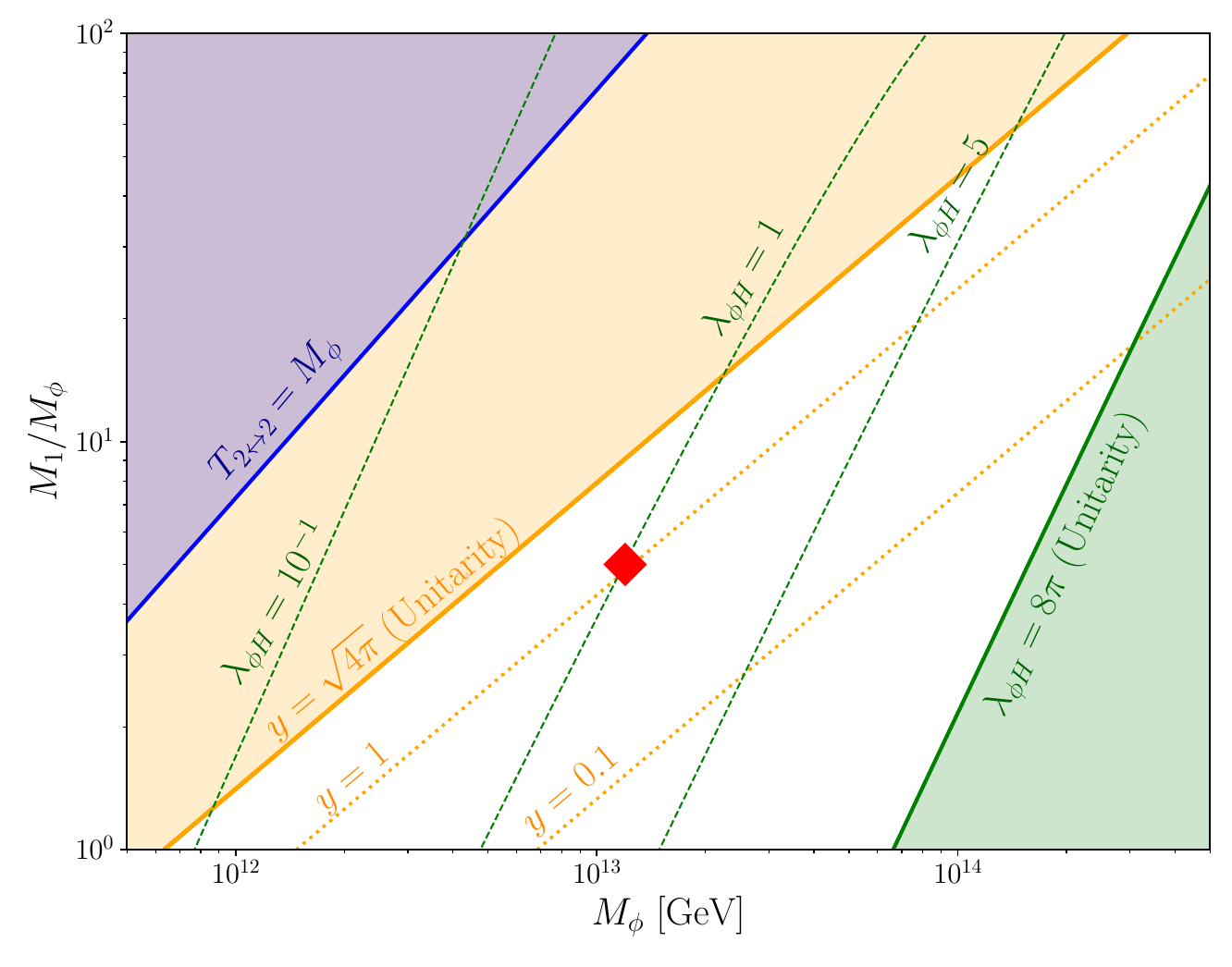}
    \caption{Regions of the parameter space leading to the observed baryon asymmetry from the decay $\phi\rightarrow LH LH$, assuming radiation domination in the early Universe. The yellow and green areas limit respectively limit the regions where the Yukawa coupling $y$ and the Higgs portal coupling $\lambda_{\phi H}$ remain perturbative, while the blue area limits the region where the scalar $\phi$ does not thermalize via scatterings with the Higgs ({\it i.e.} $M_\phi > T_{2\leftrightarrow2}$). The plot also shows contour lines of the Yukawa coupling and Higgs portal coupling. The red dice is the benchmark for model parameters as used in Figure \ref{fig:leptogenesis_thermal}.}    \label{fig:model_parameters}
\end{figure}

As a result, the available  parameter space of the model is solely controlled by the off-shell suppression factor given by the mass hierarchy $(M_1/M_\phi)^2$, which simultaneously governs the CP asymmetry and the scalar decay rate. Once again, thermal freeze-out realization favors relatively small mass hierarchies, 
\begin{equation}
    y\lesssim\sqrt{4\pi}\implies \frac{M_1}{M_\phi}\lesssim  5\times \left(\frac{M_\phi}{5\times 10^{12}\, \rm GeV}\right)^{3/4}\,
    ,
\end{equation}
as larger hierarchies rapidly push the required Yukawa coupling to too large values above the perturbative unitarity limit. At the same time, one should require that the hierarchies of temperatures $M_1> T_{\rm max}>M_\phi$ and $T_{2\leftrightarrow2}>M_\phi$ hold, to allow for production and thermalization of the scalar field, but ensure the absence of on-shell right-handed neutrino. We summarize such model parameter constraints in Figure \ref{fig:model_parameters}, assuming that the early Universe temperature satisfies the constraint $M_1>T_{\rm max}>M_\phi$.

\section{Conclusions}
\label{sec:conclusions}

We have proposed a new realization of the leptogenesis mechanism in which the baryon asymmetry of the Universe is generated even when the right-handed neutrinos are never produced in the Universe, neither by thermal nor by non-thermal processes. The model requires a scalar $\phi$ lighter than the right-handed neutrinos, which couples to them via a Yukawa coupling. This interaction allows the decay of $\phi$ into two lepton doublets and two Higgs doublets through off-shell right-handed neutrino exchange, producing a CP asymmetry from the interference between tree-level and one-loop diagrams in the four-body decay of the scalar field. The CP asymmetry in turn generates a lepton asymmetry which is later partially converted into the observed baryon asymmetry through sphaleron processes, practically without wash-out. 

We then studied two cosmological realizations of the mechanism. First, we considered the case where the scalar field $\phi$ plays the role of the inflaton and reheats the Universe through its decays. In this scenario, the baryon asymmetry is produced during reheating, and we identified regions of parameter space where the correct asymmetry can be generated while keeping the right-handed neutrinos sufficiently heavy so that they never populate the Universe. Second, we analyzed the case where $\phi$ is a generic scalar field in thermal equilibrium with the plasma. We derived the Boltzmann equations describing the evolution of the asymmetry and showed that successful leptogenesis can also occur in this setup, although for a narrower range of parameters. Indeed, the thermal history must allow $\phi$ to be produced and thermalized while keeping the right-handed neutrinos off shell, which requires the relevant temperatures to lie between the scalar and right-handed-neutrino mass scales. We have found that, in both cosmological realizations, successful baryogenesis requires a moderate hierarchy between the right-handed neutrino and scalar field masses, as a too large hierarchy suppresses both the CP asymmetry and the decay rate, and does not reproduce the observed baryon asymmetry.

Our results demonstrate that successful leptogenesis can take place in the type-I seesaw mechanism, even when the maximal temperature of the Universe is below the right-handed neutrino mass scale, and right-handed neutrinos are never produced on-shell. Extensions of the model to other seesaw frameworks will be presented in a forthcoming publication.

\acknowledgments

This work was supported by the Collaborative Research Center SFB1258 and by the Deutsche Forschungsgemeinschaft (DFG, German Research Foundation) under Germany’s Excellence Strategy - EXC-2094 - 390783311. SC was supported by the Humboldt foundation with a Humboldt Research Fellowship.

\appendix

\section{Decay rate and CP asymmetry}
\label{app:CP}
We consider the four-body decay \(\phi\to L_\ell^{\alpha_1}\,H^{\beta_1}\,L_{\ell'}^{\alpha_2}\,H^{\beta_2}\), where $\ell$ and $\ell'$ denote leptonic flavors, while $\alpha_{1,2}$, $\beta_{1,2}$ are $SU(2)_L$ indices. We compute the amplitudes under the assumption that the right-handed neutrinos are much heavier than the scalar $\phi$: $M_q^2 \gg M_\phi^2 \geq s_{ij} $, $(q= 1,..., N)$, where  $j=1,2,3,4$, $i<j$, and $s_{ij}$ the Mandelstam variables for $1 \rightarrow 4$ decay (defined as $s_{ij}=(p_i+p_j)^2$, with $p_i$ and $p_j$ the 4-momenta of any two final state particles).

The decay rate of $\phi$ in this channel reads
\begin{equation}
     \Gamma_{\text{tot}}\simeq \dfrac{1}{2 M_\phi} \sum_{\ell,\ell'}\int \overline{|\mathcal{M}^{\ell,\ell'}_\text{tree}|^2} d\phi_4\,
     ,
\end{equation}
with $d\phi_4$ the 4-body Lorentz invariance phase space, while the spin– and index–summed tree-level squared amplitude  reads,
\begin{equation}\label{eq:SquaredTree}
  \sum_{\alpha_1, \alpha_2, \beta_1, \beta_2} \sum_{\ell, \ell'}  |\overline{M_{\alpha_1, \alpha_2, \beta_1, \beta_2}^{\ell, \ell'}|^2}= 12 s_{12} \displaystyle{\sum_{i,i', q,q'}\dfrac{\mathcal{H}_{qi}y_{ii'}\mathcal{H}^*_{i'q'}  y_{q'q}^*}{ M_q M_{q'} M_i M_{i'}}}\,
  .
\end{equation}
For the evaluation of the 4-body Lorentz invariant phase space, we use the standard techniques described in \cite{ParticleDataGroup:2024cfk}, and we find,
    \begin{align}
         \Gamma_{\text{tot}}&\simeq
         \frac{M_\phi^5}{196608\,\pi^5}\displaystyle{\sum_{i,i', q,q'}\dfrac{\mathcal{H}_{qi}y_{ii'}\mathcal{H}^*_{i'q'}  y_{q'q}^*}{ M_q M_{q'} M_i M_{i'}}}\,
         ,
    \end{align}
where we used that
\begin{equation}
    \int d\phi_4\,  s_{12}\;=\; \frac{M_\phi^6}{294912\,\pi^5}\,
.
\end{equation}

If the scalar field $\phi$ is exclusively interacting with the lightest right-handed neutrinos $N_1$, then we can write down the decay rate as
\begin{align}
  \Gamma_{\text{tot}}&\simeq 
          \frac{M_\phi^5}{196608\,\pi^5}\dfrac{|\mathcal{H}_{11}|^2 |y_{11}|^2 }{M_1^4} =  \frac{ |y_{11}|^2}{196608\,\pi^5}\dfrac{\tilde m_1^2 M_\phi^3 }{ v^4} \left(\dfrac{M_\phi}{M_1}\right)^2
          \label{eq:decay_rate}\,.
\end{align}
In the main text, we assume for simplicity that $\tilde{m}_1\sim m_3$.\\

The CP asymmetry originates from the interference of the tree diagram with the absorptive (on-shell) parts of one-loop corrections. The imaginary parts arise from unitary (Cutkosky) cuts through intermediate massless $L$ and $H$ lines in the vertex, self-energy (wave-function), and box topologies, shown in Fig.~\ref{fig:all_diagrams}, and are labeled as ``vertex", ``wave function" (WF), ``scalar-contact topology," and ``box". 

For the vertex diagrams, the difference in the amplitudes between the process $\phi\rightarrow LH LH$ and its CP conjugated reads,
\begin{equation}
\begin{aligned}
   \Delta \overline{|{\cal M}|^2}\Big|_{\rm Vertex}\simeq   -\dfrac{ 3s_{12}(M_\phi^2-s_{12}-s_{34})}{4\pi }\displaystyle{\sum_{i, i', q,q',k}} \dfrac{{\rm Im}\left[\mathcal{H}_{qi} y_{ii'}  \mathcal{H^*}_{i'k}  \mathcal{H}_{k q'} y_{q'q}^* \right]}{ M_k M_q^2 M_{q'} M_i M_{i'}}\,
   .
\end{aligned}
\end{equation}
Using that 
\begin{equation}\label{eq:appendixint}
\int d\phi_4\, s_{12}(M_\phi^2-s_{12}-s_{34})\;=\;\frac{M_\phi^8}{2949120 \pi^5}\,,
\end{equation}
we obtain that the contribution to the total CP asymmetry from the vertex diagrams is,
\begin{equation}
\begin{aligned}
    \label{eq:eps-start}
\epsilon\big|_{\mathrm{Vertex}}
&\simeq
-\frac{M_\phi^2}{320 \pi}\dfrac{\displaystyle{\sum_{i, i', q,q',k}} \dfrac{{\rm Im}\left[\mathcal{H}_{qi} y_{ii'}  \mathcal{H^*}_{i'k}  \mathcal{H}_{k q'} y_{q'q}^* \right]}{ M_k M_q^2 M_{q'} M_i M_{i'}}}{\displaystyle{\sum_{i,i', q,q'}\dfrac{\left[\mathcal{H}_{qi}y_{ii'}\mathcal{H}_{i'q'}^*  y_{q'q}^*\right]}{ M_q M_{q'} M_i M_{i'}}}}\,
.
\end{aligned}
\end{equation}

Similarly, for the self-energy contribution, we find that
\begin{equation}
\begin{aligned}
   \Delta \overline{|{\cal M}|^2}\Big|_{\rm WF} \simeq &-\dfrac{ 3 s_{12}(M_\phi^2-s_{12}-s_{34})}{2\pi}\\
   &\times \displaystyle{\sum_{i, i', q,q',k}} \dfrac{ M_k
 {\rm Im}\left[\mathcal{H}_{qi} y_{ii'}  \mathcal{H^*}_{i'k}  \mathcal{H}_{k q'} y_{q'q}^* \right] + M_q {\rm Im}\left[\mathcal{H}_{qi} y_{ii'}  \mathcal{H}_{i'k}  \mathcal{H}_{k q'} y_{q'q}^* \right] }{ M_k^2 M_q^2 M_{q'} M_i M_{i'}}.
\end{aligned}
\end{equation}
It is important to note that the combination $\mathcal{H}_{qi} y_{ii'}  \mathcal{H}_{i'k}  \mathcal{H}_{k q'} y_{q'q}^*$ is real, so that it does not contribute to the asymmetry. Finally, using eq.~(\ref{eq:appendixint}) we find
\begin{equation}
\label{eq:eps-start-WF}
\begin{aligned}
\epsilon\big|_{\mathrm{WF}}
&\simeq
-\frac{M_\phi^2}{160\pi}
\dfrac{\displaystyle{\sum_{i, i', q,q',k}} \dfrac{{\rm Im}\left[\mathcal{H}_{qi} y_{ii'}  \mathcal{H^*}_{i'k}  \mathcal{H}_{k q'} y_{q'q}^* \right]}{ M_k M_q^2 M_{q'} M_i M_{i'}}}{\displaystyle{\sum_{i,i', q,q'}\dfrac{\mathcal{H}_{qi}y_{ii'}\mathcal{H^*}_{i'q'}  y_{q'q}^*}{ M_q M_{q'} M_i M_{i'}}}},
\end{aligned}
\end{equation}
which is twice as large as the vertex correction, mirroring the result of the standard leptogenesis scenario from the decay of an on-shell right-handed neutrino.

The third contribution is labeled as ``scalar-contact topology" in Figure \ref{fig:all_diagrams}. The asymmetry generated by this diagram is $\propto {\rm Im}\left[\mathcal{H}_{qi} y_{ii'}  \mathcal{H}_{i'k}  \mathcal{H}_{k q'} y_{q'q}^* \right]=0$.

Lastly, the for the ``box" topology, the coupling structure is the same as the tree-level diagram multiplied by the four-Higgs coupling constant $\lambda$. $\lambda$ is a real coupling constant, and the loop amplitude has no
absorptive part; the interference with the tree amplitude is purely real and therefore does not contribute either to the CP asymmetry.  

The total CP asymmetry is then
\begin{equation}
\begin{aligned}
    \label{eq:eps-start-tot1}
\epsilon_{\mathrm{tot}}
&\simeq -\frac{3M_\phi^2}{ 320\pi}
\dfrac{\displaystyle{\sum_{i, i', q,q',k}} \dfrac{{\rm Im}\left[\mathcal{H}_{qi} y_{ii'}  \mathcal{H^*}_{i'k}  \mathcal{H}_{k q'} y_{q'q}^* \right]}{ M_k M_q^2 M_{q'} M_i M_{i'}}}{\displaystyle{\sum_{i,i', q,q'}\dfrac{\mathcal{H}_{qi}y_{ii'}\mathcal{H^*}_{i'q'}  y_{q'q}^*}{ M_q M_{q'} M_i M_{i'}}}}.
\end{aligned}
\end{equation}
In the case of $N_1$ exclusivity, we can write down the CP-asymmetry in the following simple and recognizable form
\begin{equation}
\begin{aligned}
    \label{eq:eps-start-tot2}
\epsilon_{\mathrm{tot}}
&\simeq -\frac{3M_\phi^2}{ 320\pi}
\dfrac{\displaystyle{\sum_{k}} \dfrac{{\rm Im}\left[\mathcal{H}_{11} y_{11}  \mathcal{H^*}_{1k}  \mathcal{H}_{k 1} y_{11}^* \right]}{ M_k M_1^5}}{\displaystyle{\dfrac{\mathcal{H}_{11}y_{11}\mathcal{H^*}_{11}  y_{11}^*}{ M_1^4}}}\,
.
\end{aligned}
\end{equation}

\bibliographystyle{JHEP}
\bibliography{biblio.bib}

\end{document}